\documentclass[a4paper,fleqn,usenatbib]{mnras}

\usepackage{newtxtext,newtxmath}

\usepackage[T1]{fontenc}
\usepackage{ae,aecompl}
\usepackage{hyperref}

\usepackage{graphicx}	
\usepackage{amsmath}	
\usepackage{amssymb}	

\title[VBBinaryLensing]{\texttt{VBBinaryLensing}: a public package for microlensing light curve computation}

\author[V. Bozza, E. Bachelet, F. Bartoli\'c, T. Heintz, A. Hoag, M. Hundertmark]{
Valerio Bozza$^{1,2}$\thanks{E-mail: valboz@sa.infn.it},Etienne Bachelet$^{3}$, Fran Bartoli\'c$^{4}$, \newauthor Tyler M. Heintz$^{5}$, Ava R. Hoag$^{5}$, Markus Hundertmark$^{6}$
\\
$^{1}$Dipartimento di Fisica ``E.R. Caianiello'', Universit\`a di Salerno, Via Giovanni Paolo II 132, Fisciano, 84084, Italy.  \\
$^{2}$Istituto Nazionale di Fisica Nucleare, Sezione di Napoli, Via Cintia, 80126, Napoli, Italy.
\\
$^{3}$
Las Cumbres Observatory, 6740 Cortona Drive, Suite 102, Goleta, CA 93117 USA.
\\
$^{4}$
Centre for Exoplanet Science, SUPA, School of Physics and Astronomy, University of St. Andrews, St. Andrews KY16 9SS, UK.
\\
$^{5}$
Department of Physics, Westminster College, New Wilmington, PA 16172, USA.
\\
$^{6}$Astronomisches Rechen-Institut, Zentrum f{\"u}r Astronomie der Universit{\"a}t Heidelberg (ZAH), 69120 Heidelberg, Germany.
}

\date{Accepted XXX. Received YYY; in original form ZZZ}

\pubyear{2017}

\begin{document}
\label{firstpage}
\pagerange{\pageref{firstpage}--\pageref{lastpage}}
\maketitle

\begin{abstract}
The computation of microlensing light curves represents a bottleneck for the modeling of planetary events, making broad searches in the vast parameter space of microlensing extremely time-consuming. The release of the first version of \texttt{VBBinaryLensing} (based on the advanced contour integration method presented in \citet{Bozza10}) has represented a considerable advance in the field, with the birth of several analysis platforms running on this code. Here we present the version 2.0 of \texttt{VBBinaryLensing}, which contains several upgrades with respect to the first version, including new decision trees that introduce important optimizations in the calculations.
\end{abstract}




\section{Introduction}

Microlensing has become one of the main methods for detecting extrasolar planets \citep{GouldVietri}. After the Kepler harvest \citep{Kepler}, we have now a pretty large statistics on planets close to their parent stars, with periods below one year\footnote{See online catalogues: \url{http://exoplanets.eu} \citep{ExoEnc}, \url{https://exoplanetarchive.ipac.caltech.edu} \citep{NASA}.} \citep{Christiansen}. Radial velocities also enabled the discoveries of some gas giants further away and enriched the table of habitable worlds. Direct imaging has unveiled giant planets in the outer regions of young systems \citep{Detection}. At present, microlensing stands as the only way to explore the colder disk beyond the snow line, where most of the planetary embryos are believed to form through accretion of volatiles 	\citep{Alibert}.

Yet, microlensing events only probe the magnification maps generated by the lensing systems along a single line: the actual source trajectory. Different lens configurations may justify the same observed light curve, inducing severe discrete and continuous degeneracies in the parameters \citep{GriSaf}. Furthermore, a sparse sampling may complicate the problem with accidental degeneracies due to gaps in the coverage \citep{Degeneracies}. For all these reasons, the full exploration of the parameter space is exceptionally complicated compared to other astrophysical problems and requires the calculation of thousands of models in order to probe far apart minima in such a vast parameter space.

In addition to this, the calculation of each individual light curve takes a long computational time, because the magnification calculation has to be repeated at the epoch of each observation. If thousands of observations have been collected, we have to make thousands of magnification calculations just to evaluate an individual model.

More in detail, for any position of the source along its trajectory, in order to obtain its magnification, we need to draw the corresponding images generated by the lens and estimate their extension. For point-sources, this is relatively easy, since we can recast the lens equation into a complex polynomial of order $n^2+1$ ($n$ is the number of lenses), which can be solved numerically for any given source position \citep{Witt90}.

However, physical sources are stars with their own angular size, which cannot be neglected especially in regions where the magnification varies very quickly or is discontinuous \citep{Griest,NemWic,WitMao}. The algorithms solving this problem can be roughly distinguished in two classes: inverse ray shooting and contour integration.

In inverse ray shooting \citep{KayRefSta,Wam90,BenRhi,Ben10}, light rays are shot back from the observer to the lens plane. Only those rays landing within the source disk are counted for the magnification. In this case, we do not have to invert the lens map. A whole magnification map on the source plane can be constructed by shooting rays on a fine grid on the lens plane. Such maps can be re-used for different source positions.

In contour integration \citep{SchKay,Dom95,GouGau,Dom98}, the lens map is inverted on a collection of points on the source boundary. In this way, one obtains a corresponding collection of points on the boundaries of the images, from which the area of the images can be recovered by use of Green's theorem. In principle, this method converts a two-dimensional problem (area of the images) to a one-dimensional problem (a summation over points on the boundary). At least for uniform-brightness sources, this solution promises to be quite effective.

In a previous work \citep{Bozza10}, we have presented several solutions to make contour integration much more efficient. The accuracy of the summation is improved by a parabolic correction; an error estimate is introduced on each arc of the boundary; this allows the definition of an optimal sampling, where new points are introduced on sections of the boundary in which we find the largest error; limb darkening can be included by repeating the calculation on more annuli, adaptively introduced when needed. All these optimizations have been implemented in a code that has been at the core of RTModel\footnote{\url{http://www.fisica.unisa.it/GravitationAstrophysics/RTModel.htm}}, a completely automatic real-time modeling platform running on a 8-core workstation.

In the year 2016, a first version of the contour integration code has been made available on a public webpage with the name \texttt{VBBinaryLensing}\footnote{\url{http://www.fisica.unisa.it/GravitationAstrophysics/VBBinaryLensing.htm}}. After almost two years of extensive testing by the community, we are now ready to release version 2.0 of this code, including the implementation of some more optimizations and features to be described in this paper. The presentation of such ideas may be of interest to the whole microlensing community even from a purely theoretical perspective. The code is now freely available also on a GitHub repository\footnote{\url{https://github.com/valboz/VBBinaryLensing}} under the lesser GPL 3.0 license. It is possible to compile it as a simple \texttt{C++} library or to install a \texttt{Python} package with all the functionality of the \texttt{C++} code. The \texttt{README.md} file contains basic instructions while several example files illustrating the use of the library from both \texttt{Python} and \texttt{C++}.

In Section 2, we review contour integration and its optimizations at the base of \texttt{VBBinaryLensing}. In Section 3 we discuss the details of the new root solving routine. In Section 4, we compare the effects of requiring a relative precision goal vs an absolute accuracy goal. In Section 5 we present three new tests based on the quadrupole approximation and ghost images used to decide whether the point-source approximation is sufficient or a full contour integration is necessary. Section 6 illustrates the calculation of the magnification for a single lens and a finite source. In Section 7 we describe the full light curve calculations, explaining how parallax, orbital motion and microlensing from satellites are implemented in the code. Section 8 contains the conclusions. An appendix lists all methods and properties of the \texttt{VBBinaryLensing} library.

\section{Contour integration}

\texttt{VBBinaryLensing} is a code based on a new advanced contour integration algorithm \citep{Bozza10}. In this section, we briefly summarize the core ideas and refer the reader to that paper for more details. 

For a uniform brightness source, the microlensing amplification is just the ratio of the total angular area of the images to the source angular area
\begin{equation}
\mu=\frac{A_{Images}}{A_{Source}}. \label{mugen}
\end{equation}

We use standard coordinates for the lens/image plane and for the source plane normalized to the Einstein radius of the total lens mass. The binary lens equation reads
\begin{equation}
\vec y = \vec x - m_1 \frac{\vec x - \vec x_{m_1}}{\left|\vec x - \vec x_{m_1}\right|^2} - m_2 \frac{\vec x - \vec x_{m_2}}{\left|\vec x - \vec x_{m_2}\right|^2}, \label{LensEq}
\end{equation}
where $m_1=1/(1+q)$, $m_2=q/(1+q)$, and $q$ is the mass ratio. The lens positions can be chosen to be $\vec x_{m_1}=(-s,0)$, $\vec x_{m_2}=(0,0)$, so that $s$ is the lens separation and the origin of the reference frame is in the smaller mass $(q<1)$. This choice of the origin turns out to be the most convenient for the root solving routine (see Sec. \ref{Sec root solving}).

For a source centered at position $\vec y_S$ in the source plane and radius $\rho_*$, the source boundary can be sampled as
\begin{equation}
\vec y_i= \vec y_S+ \rho_* \left(\begin{array}{c}
  \cos \theta_i \\
  \sin \theta_i \\
\end{array} \right),
\end{equation}
where $\theta_0<\theta_1< \ldots < \theta_i<
\ldots < \theta_n=\theta_0$ is a collection of $n$ angles.
 
By solving the lens equation, we can determine the images of $\vec y_i$ and indicate them by $\vec x_{I,i}$. It is well known that for a binary lens there exist caustic curves in the source plane enclosing regions in which a point-source generates five images, while outside the caustics only three images are generated. The index $I$ then takes values from 1 to 3 if $\vec y_i$ is outside any caustics or 1 to 5 if it is inside a caustic.

Having identified the image points belonging to the same image boundary, we can obtain the total area of the images as

\begin{eqnarray}
&A_{Images} & = 
\sum\limits_I p_I \sum\limits_{i=0}^{n-1} A_{I,i}^{(t)}+A_{I,i}^{(p)} \label{AreaImages} \\& A_{I,i}^{(t)}& = \frac{1}{2}\vec x_{I,i} \wedge \vec x_{I,i+1} \\& A_{I,i}^{(p)}& = \frac{1}{24}\left[\left(\vec x'_{I,i}\wedge \vec
x''_{I,i}\right) +\left(\vec x'_{I,i+1}\wedge \vec
x''_{I,i+1}\right)\right]\Delta\theta^3,
\end{eqnarray}
where $p_I=\pm1$ is the parity of image $I$. The trapezium approximation $A_{I,i}^{(t)}$ replaces each arc between $\theta_i$ and $\theta_{i+1}$ by a straight line. The parabolic correction $A_{I,i}^{(p)}$ adds the area of the parabolic segment between the curve and the straight line as found by checking the local derivatives at the end points of the arc. As shown by Bozza (2010), these derivatives can be easily expressed from the lens equation and used to dramatically improve the accuracy of the contour integration without increasing the sampling. Similar expressions can be given for image boundaries starting or ending at a critical point, generated by a source whose boundary is partially inside a caustic. 

The magnification is then obtained by Eq. (\ref{mugen}) with $A_{Images}$ given by Eq. (\ref{AreaImages}) and the area of the source simply being $A_{Source}=\pi\rho*^2$ for a uniform source.

At each step it is essential to keep track of the error committed by our approximation of the contour integral by a finite sum. This is done by defining the following error estimators for each image arc
\begin{eqnarray}
&E_{I,i,1} &= \frac{1}{48}\left|\left(\vec x'_{I,i}\wedge \vec
x''_{I,i}\right) -\left(\vec x'_{I,i+1}\wedge \vec
x''_{I,i+1}\right)\right|\Delta\theta^3
\label{E1} \\
&E_{I,i,2} &= \frac{3}{2}\left| A_{I,i}^{(p)}\left(\frac{\left|\vec
x_{I,i}-\vec x_{I,i+1}\right|^2}{\Delta\theta^2 \left| \vec
x'_{I,i}\cdot \vec x'_{I,i+1} \right|} -1\right) \right|
\label{E2}\\
&E_{I,i,3}& = \frac{1}{10}\left| A_{I,i}^{(p)}\right |\Delta\theta^2
\label{E3},
\end{eqnarray}
where $\Delta\theta=\theta_{i+1}-\theta_i$. The first estimators checks that the parabolic approximation from the two end points of the same arc returns a consistent result; the second estimator checks that the distance between the two end points is consistent with the extrapolation from the local derivatives; the third one just checks that the sampling is fine enough that $A_{I,i}^{(p)}$ is a truly higher order correction to $A_{I,i}^{(t)}$ and is not exploding. For each of these estimators, there is a version for image boundaries starting or ending at a critical point.

In addition to these estimators, we have also introduced a check on ghost images, i.e. the two roots of the 5th order complex polynomial associated with the lens equation that do not correspond to physical images when the source is outside the caustic. If the distance between such ghost images has a minimum at some step $i$ in our sampling, we add an error
\begin{eqnarray}
&E_{g,i}&=(g_{i-1}-g_i)^2 \\ 
&g_i&=\left|\vec x_{g_1,i}-\vec x_{g_2,i}\right|,
\end{eqnarray}
and similarly for $E_{g,i-1}$. In this way, we caution that the tip of a cusp may be hidden in between our sampling.

The error estimators are essential for driving the sampling where it is actually needed. Starting from a basic sampling with two points, we gradually increase it by introducing a new $\theta_i$ partitioning the arc in which the error of the corresponding images is largest. The total error of the contour integration on the area of the images is taken as the sum of the errors of all arcs
\begin{equation}
E=\sum\limits_{I,i,j}E_{I,i,j}.
\end{equation} 
Then the absolute error on the magnification is simply
\begin{equation}
\delta\mu=\frac{E}{A_{Source}}. \label{delta mu}
\end{equation}

As well known, contour integration works fine with uniform sources but requires integration on concentric annuli to account for limb darkening. Let the source profile be given by
\begin{equation}
I(\rho)=\bar I f(\rho/\rho_*),
\end{equation}
where  $\bar I$ is the average surface brightness and the profile function is normalized to 1:
\begin{equation}
2 \int\limits_0^1dr r f(r)=1.
\end{equation}

Let $F(r)$ be the cumulative profile at fractional radius $r$
\begin{equation}
F(r)=2\int\limits_0^r dr' r'f(r').
\end{equation}

Then the magnification of an annulus with inner radius $r_{i-1}$ and outer radius $r_i$ is approximated by
\begin{eqnarray}
&& \tilde M_i= f_i\left[\mu_i r_{i}^2- \mu_{i-1} r_{i-1}^2\right] \\
&& f_i=\frac{F(r_i)-F(r_{i-1})}{r_{i}^2- r_{i-1}^2 }, \label{fi}
\end{eqnarray}
where $\mu_i$ is the magnification of a uniform disk of radius $r_i\rho_*$.

The total magnification is 
\begin{equation}
M=\sum\limits_i \tilde M_i.
\end{equation}

Also for limb darkening we can introduce error estimators for each annulus and use optimal sampling to partition annuli only when needed. We start by the point-source ($r=0$) and the source boundary ($r=1$). New radii are then inserted so as to divide the flux of the original annulus into two equal parts. In this way, the computational time is optimized in the best possible way.

The current version of \texttt{VBBinaryLensing} uses a linear limb darkening law
\begin{equation}
f(r)=\frac{1}{1-a/3}\left[1-a
\left(1-\sqrt{1-r^2}\right)\right],
\end{equation}
but a future extension to arbitrary profile will be straightforward.

\section{Root solving routine} \label{Sec root solving}

The lens equation (\ref{LensEq}) can be recast into a fifth order complex polynomial equation \citep{Witt90}, which can be solved by standard methods of complex algebra. In practice, all algorithms find one root at a time starting from an initial trial value. After one root is found, the polynomial is factored and the algorithm is repeated until all solutions are found.

Let $p(z)$ be our complex polynomial. If $z_k$ is the current trial for the root to be found, we can optimize it by setting
\begin{equation}
z_{k+1}=z_k-\frac{1}{G},
\end{equation}
where
\begin{equation}
G=\frac{p'(z_k)}{p(z_k)}.
\end{equation}

This is the so-called Newton algorithm, which works fine if we are close enough to the root.

A higher order algorithm was proposed by Laguerre:
\begin{equation}
z_{k+1}=z_k-\frac{n}{G\pm\sqrt{\sqrt{(n-1)(nH-G^2)}}},
\end{equation}
where $n$ is the degree of the polynomial, and
\begin{equation}
H=G^2-\frac{p''(z_k)}{p(z_k)}.
\end{equation}

The double sign is chosen so as to select the largest denominator and avoid singularities.

In the new version of \texttt{VBBinaryLensing}, we have adopted the Skowron \& Gould algorithm \citep{SkoGou}\footnote{\url{http://www.astrouw.edu.pl/~jskowron/cmplx_roots_sg/}}, which uses the Laguerre method when we are still very far from the root, so as to have a fast convergence and then switches to the Newton method, which is less demanding (it has no square roots and one derivative less), when we are close enough. The original \texttt{Fortran} code has been translated by us to \texttt{C++} and incorporated in the \texttt{VBBinaryLensing} library with adequate notice in the documentation. The \texttt{C++} translation of the original library has also been posted on the original Skowron \& Gould repository as a useful by-product of our development.

\begin{figure}
\resizebox{\hsize}{!}{\includegraphics{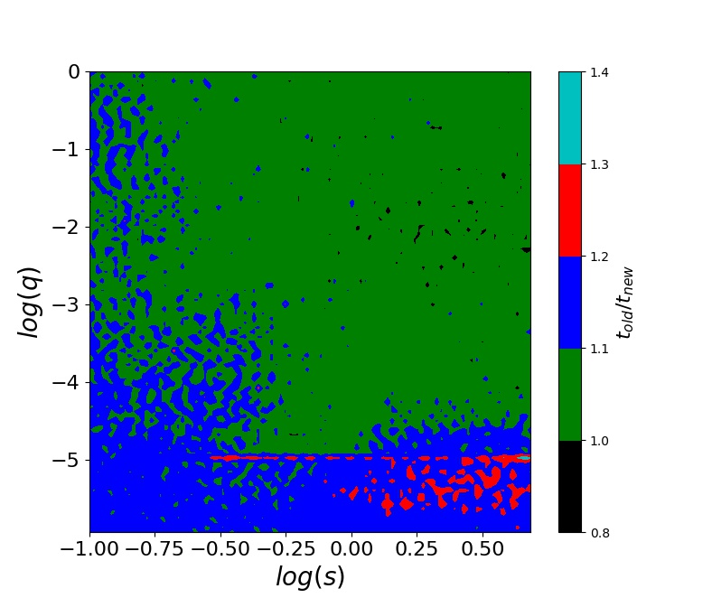}}
 \caption{Speed-up obtained on a grid of possible lensing configurations $(s,q)$. For each point we have averaged on source positions in the range $[-4,4]$ for $y_1$ and $[0,2]$ for $y_2$.}
 \label{Fig SkoGou}
\end{figure}

In Fig. \ref{Fig SkoGou} we quantify the speed-up obtained by using the Skowron \& Gould algorithm with respect to a standard Laguerre method. On average it amounts to some $10\%$. In addition, the Skowron \& Gould proves to be more robust and fail-safe against testing at extreme mass-ratios ($q\sim 10^{-9}$) .

\section{Precision vs Accuracy}

Another upgrade of \texttt{VBBinaryLensing} now implemented is the possibility for the user to specify either the absolute accuracy goal $\delta$ or a relative precision goal $\epsilon$. Then the calculation is stopped when either of the two thresholds is met. If $\delta\mu$ is the absolute error on the magnification as given by Eq. (\ref{delta mu}), the two criteria read
\begin{eqnarray}
&&\delta\mu<\delta \\
&&\frac{\delta\mu}{\mu}<\epsilon.
\end{eqnarray}

In general, when modeling real data, we need the magnification calculation to be at least as accurate as the experimental uncertainties or better. In general, photometric observations come with statistical uncertainties of the order of 1 millimag or higher, with systematics of the same order of magnitude. In most cases,  it will be sufficient to require $\epsilon=10^{-3}$ and $\delta=10^{-2}$ to match photometric errors.

\begin{figure}
\resizebox{\hsize}{!}{\includegraphics{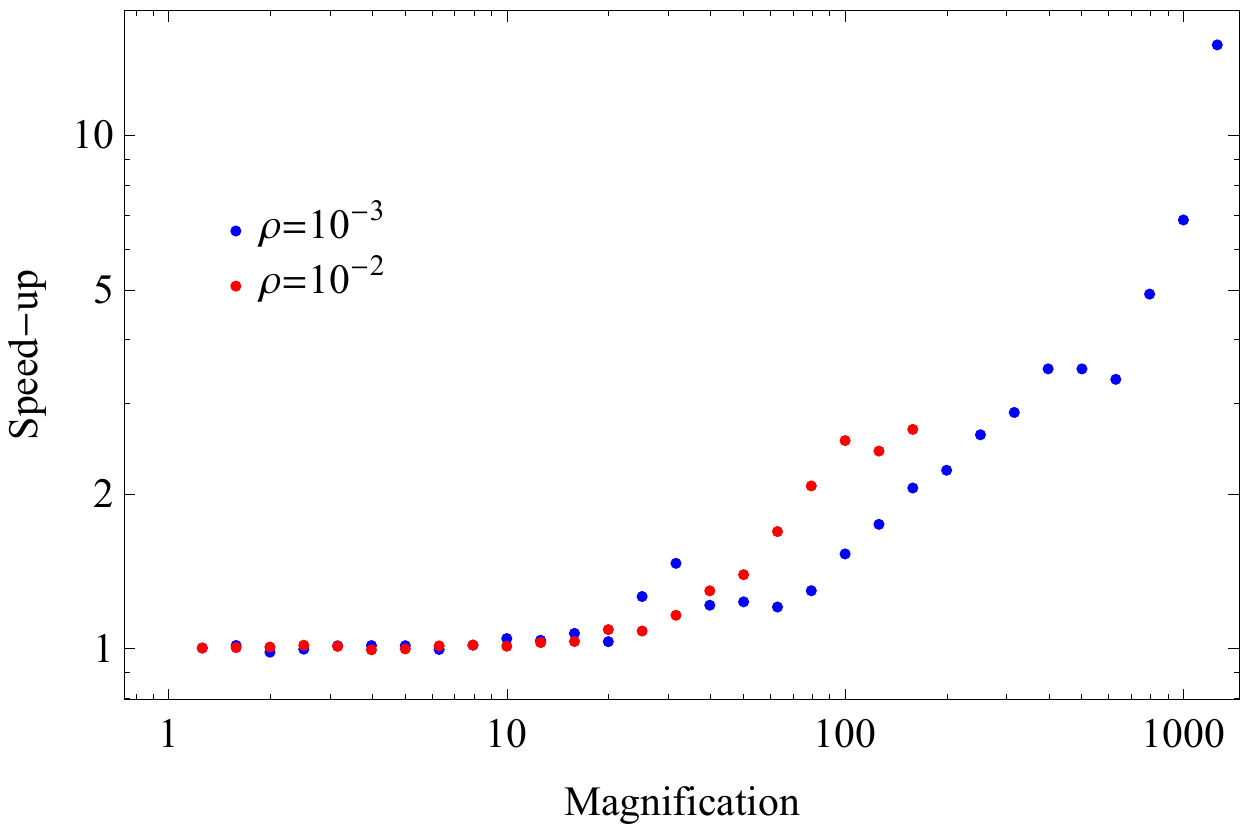}}
 \caption{Speed-up obtained using a precision goal $\epsilon=0.001$ with respect to an accuracy goal $\delta=0.01$. The lens separation is $s=1.2$ Einstein radii, the mass ratio is $q=10^{-3}$. Red points are with source radius $\rho_*=10^{-2}$, blue points with $\rho_*=10^{-3}$. Each point is an average over all entries in a grid over the source position falling within the same magnification bin.}
 \label{Fig epsdelta}
\end{figure}

With these numbers, the relative precision goal has no impact at low magnification ($\mu<10$), but significantly speeds-up the computation at higher magnification ($\mu>10$). Fig. \ref{Fig epsdelta} shows the speed-up obtained by the introduction of the relative precision goal as a function of magnification for a typical planetary lens configuration. Note that at a magnification $\mu\sim 100$ the speed-up is a factor of 3, while at magnification 1000 the speed-up may exceed 10. Saving computational time in the modeling of high-magnification events is very important, since these events are the most demanding in terms of computing time. A speed-up of few units may allow broader searches in the parameter space and more robust physical conclusions.

\section{Point-like vs finite source: a simple test to decide which way to take} \label{Sec quadrupole}

The main complication of microlensing computations indeed comes from the finiteness of physical sources. Point-like sources can be easily treated by a single inversion of the lens equation. In that case, the total magnification is just given by the sum of the absolute values of the inverse Jacobian determinant evaluated at the images positions. No sampling of the boundary, no contour integration to calculate. The algorithm summarized here in Section 2 requires a minimum of 8 points even when the source is very far from the caustics. With respect to the point-source calculation, the basic contour integration is therefore about 10 times slower.

While modeling a microlensing event, most of the survey observations fall close to the baseline. Even if some dense follow-up on the caustic crossings is present, the danger is that most of the computational time is spent where the event is least interesting, if we use the same contour integration algorithm everywhere.

However, choosing the right point where to switch from point-source to finite-source is far from straightforward and very tricky. In practice, modelers use to define an ``anomaly window'' by visual inspection of the data, and instruct their codes to use point-source outside this window and the full finite-source calculation within the anomaly window. Sometimes, some quadrupole or hexadecapole approximations are also employed in intermediate ranges around the anomaly window. 

Apparently, it is impossible to transfer this approach to automatic modeling platforms or when we aim at analyzing thousands of anomalous events at once. Furthermore, by a trivial extrapolation of the rate of anomalous events found in current ground-based surveys to the number of microlensing events expected by future space surveys such as WFIRST, we must be ready to model thousands anomalous events simultaneously. The challenge posed by such data flows requires removal of any human intervention in the identification and modeling of interesting microlensing events. In this respect, the code itself must be able to decide whether to save time by the point-source approximation or to invest on the full finite-source calculation.

Let us state the terms of the problems more precisely. For any lens configuration, parameterized by the lens separation $s$ normalized to the Einstein radius, and mass ratio $q$, we have to identify the regions of the source plane $(y_1,y_2)$ in which we need to switch from the point-source to the finite-source calculation. The boundary of these regions can be simply defined as the locus of points where the difference between the two calculation exactly matches the accuracy goal $\delta$ we wish to reach. This requirement gives an operational definition of the two regimes, their respective regions of validity and their boundary. Finally, we expect that the respective extension of the two regions depend on the source size $\rho_*$: for smaller sources we can safely use the point-source approximation up to small distances from the caustics, while for larger sources we need to use the finite-source algorithm even far away.

The condition to accept the point-source result can then be formalized by the equation
\begin{equation}
\left|\mu_{FS} - \mu_{PS} \right| < \delta, \label{Cond gen}
\end{equation}
where $\mu_{PS}$ is the point-source magnification and $\mu_{FS}$ is the finite-source magnification for a given set of parameters $(s,q,\rho_*,y_1,y_2)$. 

Our strategy is to start with the calculation of the point-source magnification $\mu_{PS}$ and use the local information made available by this calculation to predict the discrepancy between the finite-source and point-source results. In the following subsections, we introduce three tests respectively based on the quadrupole expansion, the criticality of ghost images, and the distance from planetary caustics. These three tests complement each other so as to cover the whole sensible regions defined by the inequality (\ref{Cond gen}) by their overlapping patches. These tests require very little additional computation to the pure point-source magnification but allow to save lots of computational time. Let us see them in detail.

\subsection{Quadrupole test}

It is fair to say that in any realistic calculation we will work with an accuracy goal $\delta \ll 1$. In fact, in typical modeling problems, we would range from $\delta=10^{-2}$ to $\delta = 10^{-4}$ depending on the accuracy needed to minimize numerical error in the $\chi^2$ calculation. This means that the switch from point-source to finite-source at the boundary of the inequality (\ref{Cond gen}) will occur in a perturbative regime, i.e. when the finite-source effect is just a small correction to the magnification. At this point, the first ``failure'' of the point-source approximation will be due to higher orders in a multipole expansion in $\rho_*$ exceeding $\delta$. Such multipoles have been studied by several authors \citep{PejHey,Gou08,Cas17}. Following \citet{PejHey}, the finite-source magnification for a uniform-brightness source can be expressed as
\begin{equation}
\mu_{FS}=\mu_{PS}+\frac{\rho_*^2}{8} \Delta \mu_{PS} +\frac{\rho_*^4}{192} \Delta\Delta \mu_{PS}+o(\rho_*^4), \label{muFS expansion}
\end{equation}
where $\Delta=\partial_{y_1}\partial_{y_1}+\partial_{y_2}\partial_{y_2}$ is the Laplacian operator. The second order term is called the quadrupole and the fourth order term is the hexadecapole. \citet{Gou08} suggests to calculate these Laplacians by sampling $\mu_{PS}$ on a collection of points on the source boundary and half-way from the boundary. \citet{Cas17} proposes to calculate local derivatives at the source center using the complex notation, where such derivatives can be easily expressed. Since we just want to know whether the point-source approximation is sufficient, we need a test that does not involve further lens map inversions, otherwise this test would not save any time compared to the full finite-source calculation. Therefore, we follow an approach similar to \citet{Cas17}.

The lens equation in complex notation reads
\begin{equation}
\zeta=z+f(\bar{z}),
\end{equation}
where $\zeta=y_1+i y_2$ represents the source position in complex coordinates, $z=x_1+i x_2$ is the generic point in the lens plane, the bar denotes the complex conjugation, and
\begin{equation}
f(z)\equiv-\frac{m_1}{z-s}-\frac{m_2}{z}.
\end{equation}

The Jacobian determinant of the lens equation simply reads
\begin{equation}
J=1-f'\bar{f}'.
\end{equation}
The point-source magnification at a source position $\zeta$ is obtained by summing the absolute values of the inverse Jacobian determinants calculated on each of the images $z_I$:
\begin{equation}
\mu_{PS}=\sum\limits_I \frac{1}{\left|J(z_I)\right|}.
\end{equation}

The first correction from finite-source comes from the quadrupole term in Eq. (\ref{muFS expansion}). For each image position $z_I$, the quadrupole correction in complex coordinates reads
\begin{equation}
\mu_{Q_I}=-\frac{2\mathrm{Re}\left[3\bar f'^3 f''^2-(3-3J+J^2/2)|f''|^2+J\bar f'^2 f''' \right]}{J^5} \rho_*^2, \label{Quad}
\end{equation}
where both $f$ and $J$ are evaluated at each image position $z_I$.

In the computation of this quantity, $J$ is already available from the point-source magnification calculation. The derivatives of $f(z)$ are easily calculated from its analytical expression, since it is just a rational function. The whole computation can be optimized in a few lines of code, which are good candidates to replace repeated calls to the root-solving routine, as required by the full finite-source calculation. So, as a first step, we shall definitely require that the full finite-source calculation is needed whenever the total quadrupole correction for all real images exceeds the accuracy goal $\delta$. Formally, this condition reads
\begin{equation}
\sum\limits_I c_1|\mu_{Q_I}|<\delta. \label{Cond 1}
\end{equation}
The coefficient $c_1$ is to be chosen empirically so as to ensure enough safety margin around the dangerous regions. We will set this in the final tuning of our algorithm. 

After some testing, we have soon realized that the condition (\ref{Cond 1}) is not sufficient to cover all regions for which the point-source approximation fails. This could have easily expected considering that the quadrupole is only the second order term in a power series. It may occasionally vanish while higher order corrections dominate. In particular, this happens around cusps: if we choose a specific approach angle, we may reach a cusp with zero quadrupole correction. However, we do not want to calculate arbitrary higher orders in the source size expansion, thus losing any advantage in our approach. 

\begin{figure}
\resizebox{\hsize}{!}{\includegraphics{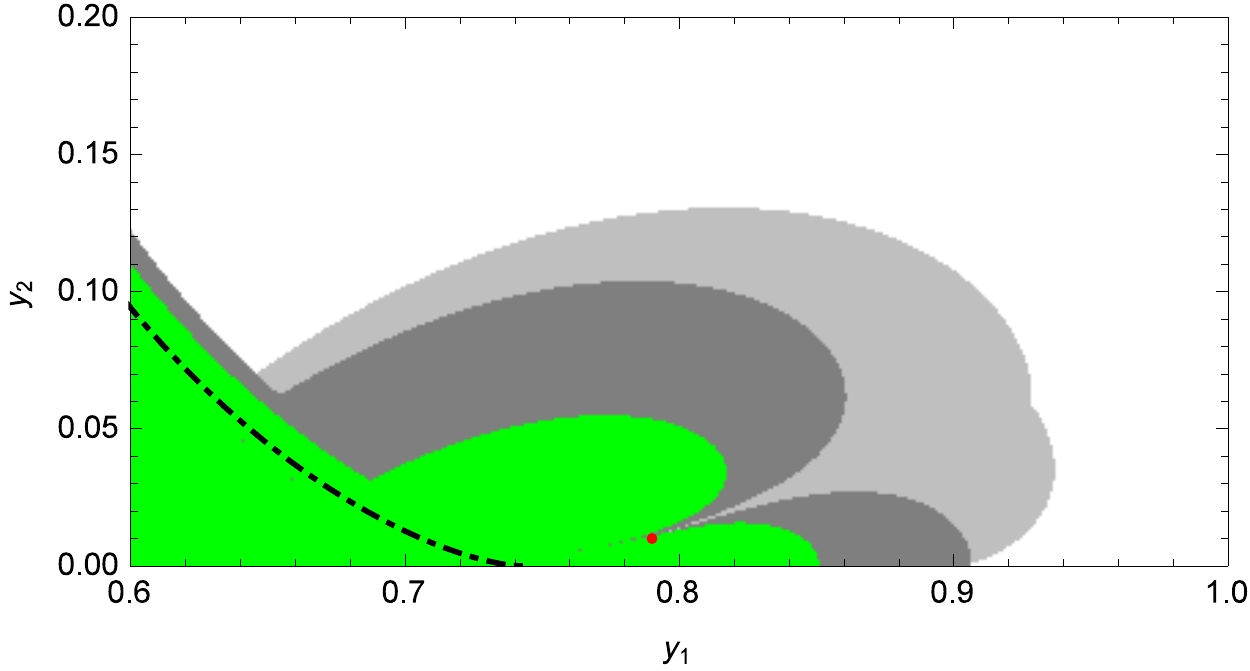}}
 \caption{Example of a map around the right cusp for a configuration with $s=1.35,q=0.32,\rho_*=10^{-2},\delta=10^{-2}$. The caustic is the dot-dashed line. In the green region, the finite-source correction is necessary, as $|\mu_{FS}-\mu_{PS}|>\delta$. This region is safely enclosed within the region covered by Eq. (\ref{Cond 1}), shown in dark grey. However, this condition leaves one red point outside, in the thin inlet where the quadrupole correction vanishes. By adding the cusp correction according to Eq. (\ref{Cond Quad}), the safety margin is extended to the light grey region, filling the inlet and solving the problem.}
 \label{Fig cusp}
\end{figure}

Cusps are defined by the vanishing of the tangential vector to the caustic. Following \citet{PetWit}, this condition can be expressed in the complex formalism by
\begin{equation}
\frac{\partial \zeta}{\partial z} H_z + \frac{\partial \zeta}{\partial \bar z} \bar H_z=0; \; \; \; H_z=2i \frac{\partial J}{\partial \bar z},
\end{equation}
which for our binary lens without shear can be recast in the form
\begin{equation}
\mathrm{Im} \left[ \bar f'^{3/2} f'' \right]=0. \label{Cusp}
\end{equation}

The complex quantity appearing here is just the square root of the first term appearing in the quadrupole correction (\ref{Quad}). Note that if Eq. (\ref{Cusp}) is fulfilled, then also the imaginary part of the squared quantity vanishes (the opposite is not true). Remembering that our purpose is to patch the quadrupole correction where it vanishes close to cusps, we look for the simplest function to add to Eq. (\ref{Cond 1}) in order to achieve this goal. This function must contain the condition (\ref{Cusp}) so as to track the cusp position correctly. It turns out that a perfect job is already achieved by the minimal correction
\begin{equation}
\mu_C= \frac{6\mathrm{Im}\left[3\bar f'^3 f''^2 \right]}{J^5} \rho_*^2.
\end{equation}  
Eq. (\ref{Cond 1}) is then updated to
\begin{equation}
\sum\limits_I c_Q\left(|\mu_{Q_I}|+|\mu_{C_I}|\right)<\delta. \label{Cond Quad}
\end{equation}

In Fig. \ref{Fig cusp} we show an example of this cusp problem solved by this correction. We see that the simple quadrupole condition (\ref{Cond 1}) defines the dark grey region, which safely encloses most of the actual points in which the finite-source effect exceeds the accuracy goal (shown in green). However, one red point is left outside by this condition. Indeed we can clearly see that the quadrupole correction vanishes along a straight line reaching the cusp by a fixed angle. The correction term we introduce in Eq. (\ref{Cond Quad}) extends the safety buffer to the light grey region, thus solving the problem.

\subsection{Ghost images test}

When the source approaches a cusp from the outside, one of the images is magnified well before the caustic crossing, thus triggering the quadrupole condition described before. However,  when approaching a fold, none of the pre-existing images are magnified. In fact, a fold crossing shows up with a discontinuous slope change in the microlensing light curves, corresponding to the sudden creation of a new pair of images. The bump starts as soon as the limb of the source touches the fold. When this happens, there is no warning of the catastrophe from the other three images. Then Eq. (\ref{Cond Quad}) only intervenes for sources on the inner side (when all 5 images participate), but is useless on the outer side.

The fold crossing is characterized by the merger of the two ghost images, which become degenerate and then separate again as true solutions of the lens equation. Since these ghost images are made available by the root solving routine, we can try to ask how far they are from being degenerate.

The Jacobian determinant is a continuous function of $z$, whether we evaluate it on a real image or not. At the fold crossing the two ghost images coincide $z_{G_1}=z_{G_2}$ and $J(z_{G_{1,2}})=0$. If our source center is just outside the fold, the two ghost images of the source center (which are available from the point-source calculation) will still be distinct but $J(z_{G_{1,2}})$ will be close to zero. A local expansion of $J$ gives (we use $z_G$ for either of the two ghost images to shorten notation)
\begin{equation}
J(z)=J(z_{G})+\frac{\partial J}{\partial z}dz+\frac{\partial J}{\partial \bar z}d \bar z. \label{Delta J}
\end{equation}
If we move from the source center by a quantity $d\zeta= \rho_* e^{i\phi}$, we must find the corresponding shift in the ghost images. These satisfy the equation
\begin{equation}
\zeta=z_G+f(\tilde z), \; \; \tilde z \equiv \bar \zeta-f(z_G).
\end{equation}
Note that real images satisfy the same equation with the additional constraint that $\tilde z =\bar z$. For ghost images, this does not hold and we have to pair this equation with its conjugate
\begin{equation}
\bar \zeta=\bar z_G+f(\bar {\tilde z}), \; \; \bar {\tilde z } \equiv \zeta-f(\bar z_G).
\end{equation}

\begin{figure}
\resizebox{\hsize}{!}{\includegraphics{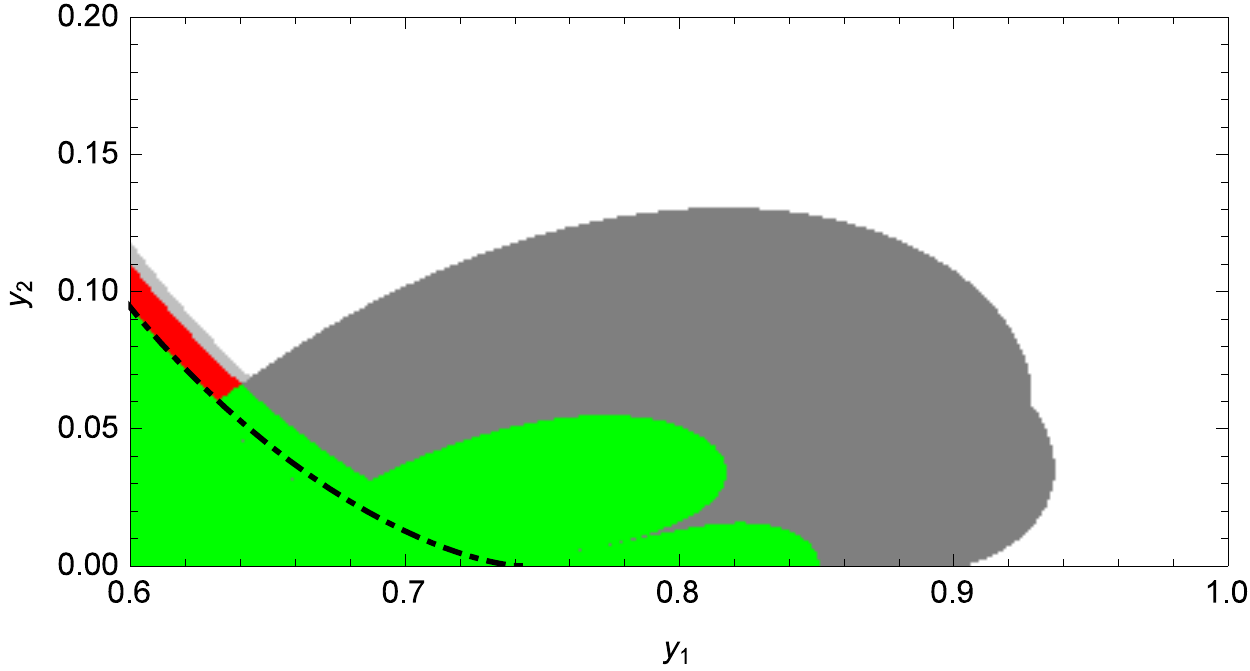}}
 \caption{Same configuration as Fig. \ref{Fig cusp}. Here we have the zone covered by the quadrupole test in dark gray. This would leave out a red stripe along the fold where the limb of the source touches the fold. This stripe is safely enclosed by the ghost images test (\ref{Cond Ghost}), reaching out to the light gray stripe.}
 \label{Fig fold}
\end{figure}

\begin{figure}
\resizebox{\hsize}{!}{\includegraphics{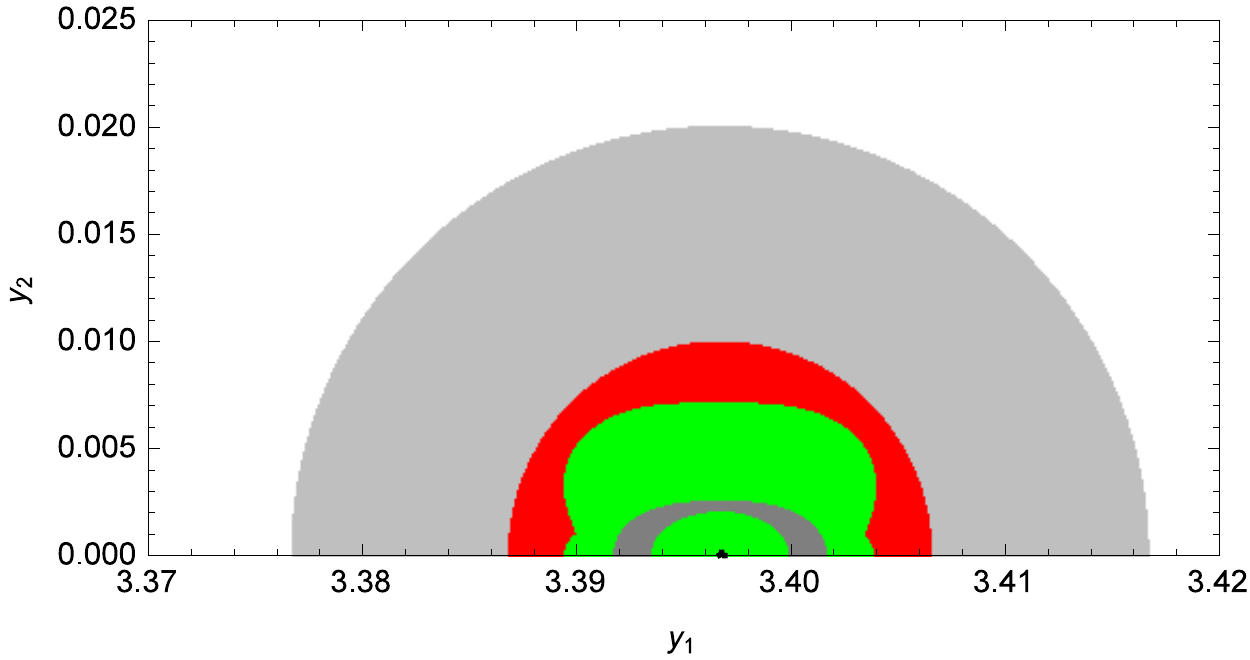}}
 \caption{Map around the planetary caustic for the configuration $s=3.67, \; q=10^{-6}, \;\rho_*=10^{-2}, \; \delta=10^{-2}$. The finite-source is required in the green and red regions. The quadrupole and ghost images tests would trigger on the green region only, leaving out a considerable annulus around the caustic. Including the planetary test (\ref{Cond plan}), we have enough buffer around the sensible region.}
 \label{Fig plan}
\end{figure}

Differentiating the two equations, we obtain the following relation between $(d\zeta,d\bar\zeta)$ and $(dz, d \bar z)$ 

\begin{equation}
\left(\begin{array}{c}
dz \\ d \bar z
\end{array}\right)=\left( \begin{array}{cc}
\frac{\tilde J}{1-f'(\tilde z)f'(\bar {\tilde z})} & \frac{\bar {\tilde J}}{1-f'(\tilde z)f'(\bar {\tilde z})}f'(\tilde z) \\
\frac{\tilde J}{1-f'(\tilde z)f'(\bar {\tilde z})}f'(\bar {\tilde z}) & \frac{\bar {\tilde J}}{1-f'(\tilde z)f'(\bar{ \tilde z})}
\end{array} \right) \left(\begin{array}{c}
\rho_* e^{i\phi} \\ \rho_*e^{-i\phi}
\end{array} \right),
\end{equation}
where
\begin{equation}
\tilde J= 1-f'(z_G)f'(\tilde z)
\end{equation}

Note that this trivially collapses to the usual Jacobian matrix when $\tilde z =\bar z_G$.

Now we can use this relation in Eq. (\ref{Delta J}) to express the change in the Jacobian function $J$ evaluated at a ghost image when we move from the center of the source to the limb. If $J(z_{G})$ is close to zero, it is possible that $J(z)$ will vanish on some points on the limb for some values of $\phi\in[0,2\pi]$. The maximum excursion of $J(z)$ from $J(z_{G})$ will be obtaining by setting the derivative with respect to $\phi$ to zero. This gives an equation for $\phi$ that can be solved to find the most dangerous point $\phi_M$. After plugging this back in Eq. (\ref{Delta J}), we obtain an equation of the kind
\begin{equation}
J(z)=J(z_{G})\pm max(\Delta J|_{\rho_*}). \label{Delta Jmax}
\end{equation}
Both $\phi_M$ and $max(\Delta J|_{\rho_*})$ can be calculated analytically, but we do not want to weigh the treatment down by lengthy expressions. In the end, if we want to test that we are far enough from a fold, we just want to make sure that $J(z)$ never vanishes on any points in the source limb, which requires
\begin{equation}
|J(z_{G})| > |max(\Delta J|_{\rho_*})|.
\end{equation}
Putting the exact expressions in this equation, we get
\begin{equation}\frac{1}{2}
\left|  J(z_{G})\frac{\tilde J^2}{\tilde J f''(\bar z_G)f'(z_G)- \bar {\tilde J}f''(z_G)f'(\bar z_G)f'(\tilde z) }  \right| > c_G\rho_*. \label{Cond Ghost}
\end{equation}
The left hand side represents the estimated distance on the source plane needed to send the ghost image to a critical point. If it is larger than the source size, we are sufficiently far from the fold, otherwise we have to switch to the finite-source calculation. The coefficient $c_G$ is set empirically from test maps in order to leave enough safety buffer, as for $c_Q$ in the case of the quadrupole test. Since we have two ghost images, both of them have to pass this test separately.

The implementation of this test requires the calculation of $\tilde z$ and the functions $f(\tilde z)$ and $\tilde J$. With respect to the quadrupole test on the real images, we do not need $f'''$, so the two tests are still of comparable complexity.

\begin{figure*}
\resizebox{\hsize}{!}{\includegraphics{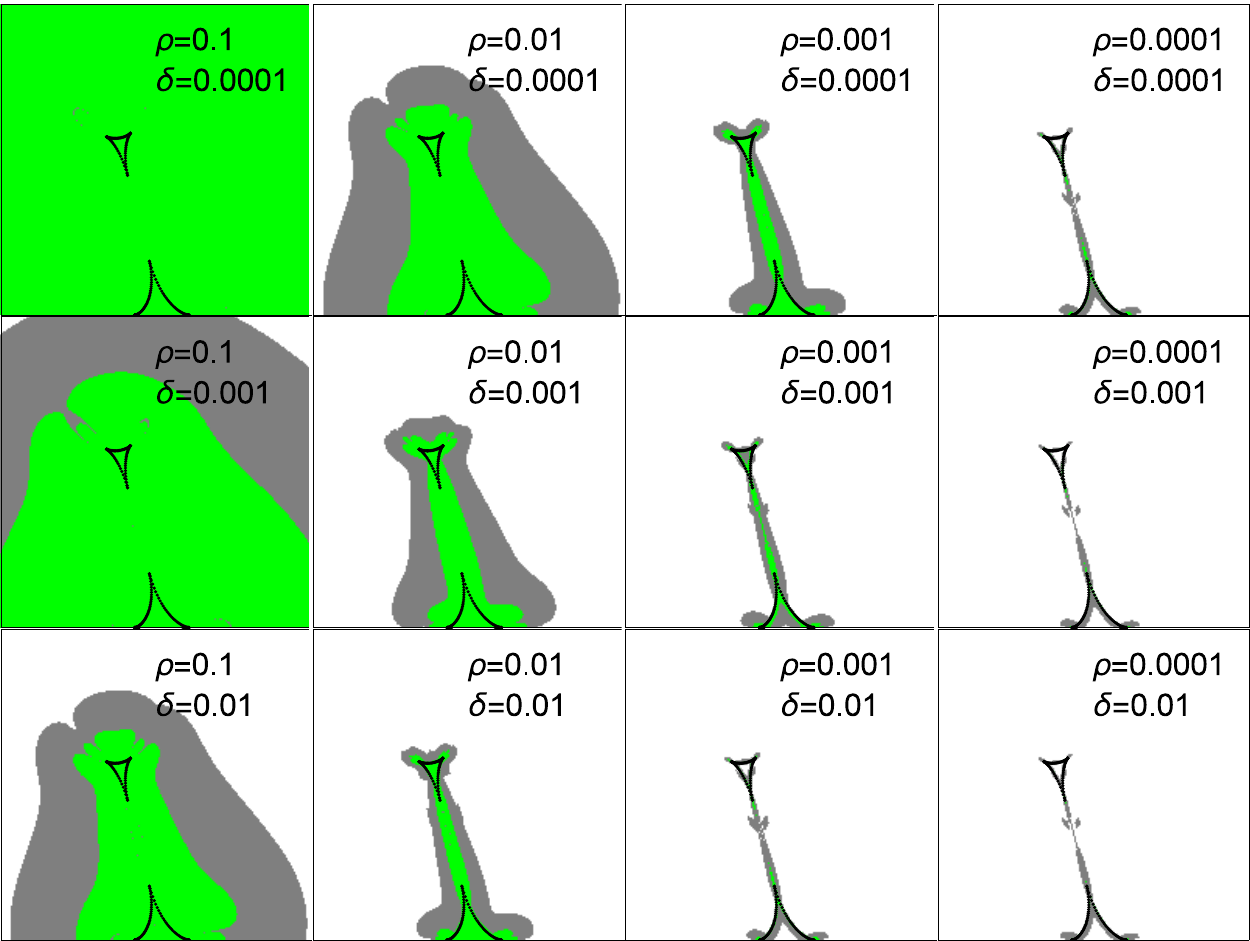}}
 \caption{Maps built for $s=0.67$ and $q=0.56$ with the indicated values of the source radius $\rho$ and the accuracy goal $\delta$. Caustics are shown in black. The green regions are those in which the difference between the finite-source and point-source magnification exceeds $\delta$. These regions are well-enclosed within the gray regions, defined by the three conditions described in this section.}
 \label{Fig maps}
\end{figure*}

\subsection{Planetary test}

The ghost images test presented in the previous section works excellently for large binary caustics, but may fail for planetary caustics when the source is so large that its limb starts to touch the planetary caustic when its center is still so far that the local gradient of the Jacobian is insensitive to the presence of the planet and is actually dominated by the primary component. This kind of failure, however, can be easily patched by checking that the distance from the center of the source to the center of the planetary caustic is greater than the source radius. In practice, the condition is

\begin{equation}
|\zeta-\zeta_{pc}|^2>c_P (\rho_*^2+\Delta_{pc}^2), \label{Cond plan}
\end{equation}
where $c_P$ is the usual coefficient to be determined empirically by the maps. The position of the planetary caustic in our frame centered on the planet is \citep{GriSaf}
\begin{equation}
\zeta_{pc}= -\frac{1}{s},
\end{equation}
and we have added a term $\Delta_{pc}$ in quadrature to the source radius to account for the extension of the caustic. The semi-extension of the caustic \citep{Bozza99,Bozza00,Dom99,Han06} is $2q^{1/2}/s^2$ for  wide planets ($s>1$) and  $2q^{1/2}/s$ for close planets ($s<1$, here we take the semi-distance between the two triangular caustics). Then we define
\begin{equation}
\Delta_{pc}= 3\frac{q^{1/2}}{s},
\end{equation}
which nicely tracks the extension of the planetary caustics in the close regime and is even too generous in the wide regime. The coefficient 3 allows enough buffer around the caustics. Fig. \ref{Fig plan} illustrates how the test succeeds for very small planetary caustics. We finally note that the planetary test is only needed for $q<0.01$.

\subsection{Broad testing on fine maps}

Up to now, we have defined three test that only use local information at the center of the source to cover all possible cases in which we need to switch to the full finite-source calculation. For each test we can still play with the coefficients $c_Q,c_G,c_P$ to decide how much buffer we want around the dangerous regions. Of course, it is recommendable to avoid switching at the very last time from point-source to finite-source, because this would generate a discontinuity in the output at the accuracy level $\delta$. Nevertheless, we definitely want to contain the waste of time in safe areas of the source plane. The compromise proposed in our code is
\begin{eqnarray}
&& c_Q=6 \\
&& c_G=2 \\
&& c_P=2. 
\end{eqnarray}

As can be imagined, the three conditions presented here are the output of an extensive testing with fine maps of the source plane for a wide range of  parameters. We have taken $q \in [10^-6,1]$, $s \in [0.1,4]$, $\rho_* \in [10^{-4},10^{-1}]$, $\delta \in [10^{-4},10^{-2}]$ and $y_1 \in [-3,4]$, $y_2 \in [0,2] $ in steps of $10^{-2}$ or $10^{-3}$ for $q<0.1$ (planetary case). These maps have allowed us to identify all possible cases and tune the coefficients so as to achieve the best performance and robustness. Since smaller sources might require an even finer grid for testing, which would be unpractical, we have replaced each $\rho_*$ appearing in conditions (\ref{Cond Quad}) and (\ref{Cond Ghost}) by $\rho_*+10^{-3}$. In practice, we do not shrink the buffer below that of a source with radius $10^{-3}$. Indeed such buffer is already so tiny that it does not make too much sense to decrease it further.

To make an example of the test maps we have built, in Fig. \ref{Fig maps} we show a table of maps obtained for the same lens configuration with various sizes of the source and accuracy goals. We note that the at lower accuracies and smaller sources the point-source approximation works excellently down to the caustic, while larger sources and finer accuracies require an earlier switch from point-source to finite-source. In any case, the three conditions defined in this section perfectly enclose any dangerous regions.

\section{Finite-source effect on single lens} \label{Sec ESPL}

We have extensively discussed the upgrades on the code for the binary lensing magnification, since this represents the highlight of the \texttt{VBBinaryLensing} library. However, the library also offers the possibility to calculate the single-lens magnification in several flavors: point-source, finite-source, finite-source with limb darkening.

Traditionally, the finite-source effect for uniform brightness disks is calculated using elliptic integrals \citep{WitMao}. The same approach can be extended to arbitrary limb darkening profiles provided the source is small enough compared to the Einstein angle \citep{Yoo04}.

After all, the calculation of elliptic integrals is not for free and requires several operations. Since the finite-source effect basically depends on the two parameters $u$ and $\rho$ (leaving apart the limb darkening for the moment), we have explored the possibility of pre-calculating the finite-source corrections for a wide enough range of these two parameters. These pre-calculated corrections have been stored in a binary table that is loaded at the first call of the finite-source-point-lens magnification function. We have then realized that reading a pre-calculated correction to the Paczynski magnification in a table is extremely faster than going through all elliptic integrals. Of course, the table must be fine enough to track the corrections with a simple linear interpolation between two consecutive entries in the table. We have chosen an accuracy goal of $10^{-3}$. Finally, to account for limb darkening, we have adopted the same algorithm as for the binary lens magnification, i.e. we calculate the magnification on multiple annuli until we match the required accuracy.

\begin{figure}
\resizebox{\hsize}{!}{\includegraphics{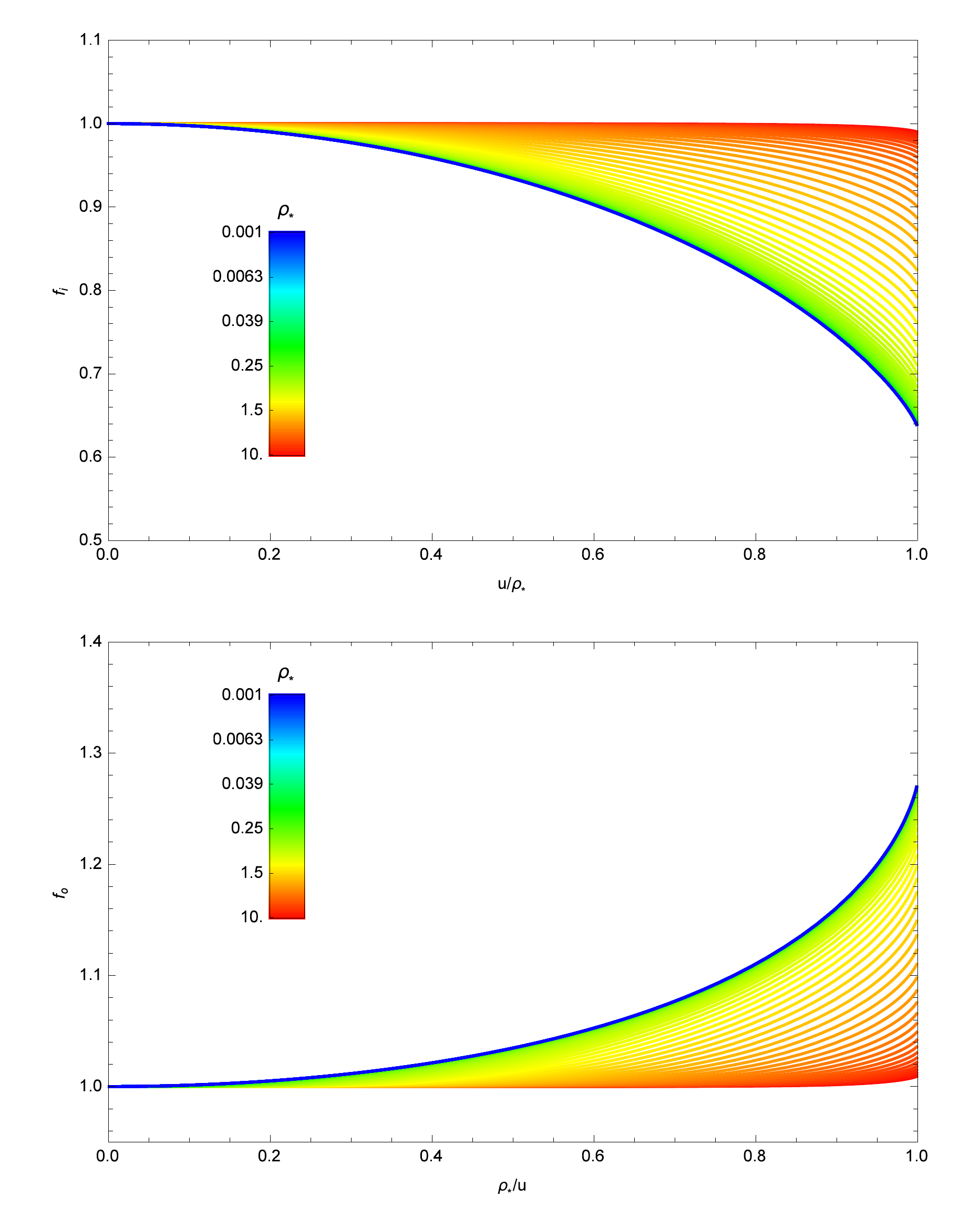}}
 \caption{Finite-source correction as a function of $u/\rho_*$ for $u<\rho_*$ (top) and $u>\rho_*$ (bottom). 101 curves are shown for source sizes in the range $[10^{-4},10^{+1}]$ with a logarithmic step. The larger the source the larger $f_o$ and the smaller $f_i$.}
 \label{Fig ESPL}
\end{figure}

More in detail, the uniform-source magnification is
\begin{equation}
\mu(u,\rho_*)=\left\{ 
\begin{array}{ll}
\frac{u^2+2}{u\sqrt{u^2+4}} f_o(\rho_*/u,\rho_*) & u>\rho_* \\
\sqrt{1+\frac{4}{\rho_*^2}} f_i(u\rho_*,\rho_*) & u<\rho_* 
\end{array} 
\right.
\end{equation}
The functions $f_o$ and $f_i$ can be read from \citet{WitMao} in terms of elliptic integrals. We show them in Fig. \ref{Fig ESPL} for a variety of source radii as functions of the ratio $u/\rho_*$ or its inverse.

Another considerable advantage in using pre-calculated tables with respect to the approach by \citet{Yoo04} is that we are not limited to small source radii $\rho_*\ll 1$. Our approach allows to investigate sources larger than the Einstein radius without any limitations. This is particularly interesting for the study of free-floating planets, which are characterized by small Einstein radii and often fall in this case.

\section{Light curve calculations}

The basic functions included in \texttt{VBBinaryLensing} allow to calculate the binary and single lens magnifications for specific lens and source configurations. However, it is also useful to have functions that calculate the magnification along an entire microlensing light curve as a function of typical fit parameters. For example, static binary lens models depend on the lens separation, mass ratio and source radius, while the source position is expressed in terms of some geometric parameters (closest approach distance, time of closest approach, trajectory angle, Einstein time).

Our code offers the possibility to calculate a single point on the light curve at a given time for any set of parameters. Alternatively, one can submit a full array of observation epochs and get back an array of magnifications, thus calculating the whole light curve with one single call.

\subsection{Parallax, satellites and orbital motion}

To a first extent, a microlensing event is short enough to approximate the relative angular motion of lens and source as rectilinear. Furthermore, the two lenses are taken as fixed. However, we know that the Earth revolves around the Sun in one year and that the two components of a binary system must orbit around their common center of mass. In general, the time-scales of these two motions are much higher than the time-scale of the microlensing event, so that a ``static'' model is sufficient to describe most binary events. However, the interest in such higher order effects is rapidly increasing thanks to the availability of more accurate photometry and sampling, allowing to track long-term effects. In addition, observations from space open the possibility to study the geometric parallax systematically for all events, granting access to this precious information for the near totality of events.

In our code it is possible to include parallax in the set of parameters. The Earth orbit is calculated using the JPL approximate ephemeris formulae\footnote{\url{https://ssd.jpl.nasa.gov/txt/aprx_pos_planets.pdf}}, which is simple and largely sufficient for microlensing parallax purposes. The user can choose between the North-East reference frame to express the components of the parallax vector and the parallel-perpendicular frame, where the reference vector in this case is the projection of the Earth-acceleration on the sky. Furthermore, it is possible to specify the reference time $t_{0,par}$, which can coincide with the closest approach time $t_0$ or can be separately specified. Then all deviations from the straight motion of the Earth are calculated with respect to the chosen $t_{0,par}$ (see e.g \citet{Skowron11}). Finally, if one wants to calculate the light curve as observed by a satellite, it is possible to use a position table of the spacecraft in the JPL format.

The orbital motion of a binary lens may become important if the caustics are relatively large (as for intermediate binaries). In this case, small changes in the separation of the lenses may translate into a sensible change in the shape of the caustics while the source is still crossing them or passing close enough for the magnification to be affected. In most studies, the linear approximation is adopted, with the addition of the two parameters $ds/dt$ (tracking the change in the separation of the lenses) and $d\alpha/dt$ (tracking the rotation of the system) \citep{Albrow00}. This approximation becomes unphysical if followed too far from the reference time $t_{0,orb}$ (typically equal to $t_{0,par}$). Therefore, we prefer to propose a circular orbit for the lens as a built-in function in the library, with velocity parameters $(\gamma_\parallel,\gamma_\perp,\gamma_z)$ as described in \citet{Skowron11}. Even if the radial velocity $\gamma_z$ would be poorly constrained in many cases, in our opinion it is better to stay on the safe side. The inclusion of the option for a full Keplerian orbit should come with future versions of the code without too much trouble.

\section{Conclusions}

With large-scale ground-based surveys detecting more and more microlensing events \citep{ground}, the incoming LSST widening the microlensing window to the whole sky \citep{LSST}, and the planned WFIRST mission promising thousands of planetary events \citep{WFIRST}, the need for fast data processing is becoming more and more urgent. The distinction of genuine microlensing planets from false positives has to be made as automatic as possible, minimizing the human intervention on exceptionally puzzling cases. To this purpose, a few publicly available modeling platforms for microlensing are currently being developed: \texttt{pyLIMA}\footnote{\url{https://github.com/ebachelet/pyLIMA}} \citep{pyLIMA}, \texttt{MulensModel}\footnote{\url{https://github.com/rpoleski/MulensModel}} \citep{MulensModel}, \texttt{muLAn}\footnote{\url{https://github.com/muLAn-project/muLAn}}. Although the goals of these platform differ somewhat, they all have in common the embedding of \texttt{VBBinaryLensing} as the basic tool for calculations of binary lens magnifications. The continuous development, maintenance and upgrade of our software is therefore of fundamental importance of several higher level software tools that aim at upholstering microlensing with more appealing and friendly interfaces.

In the new version presented in this paper, we have introduced several novelties for speeding-up the light curve calculation in all regimes. The Skowron \& Gould algorithm for the basic root solving routine provides a modest speed-up but more robustness. The possibility to specify a precision goal dramatically speeds-up high-magnification calculations. Three new tests now drive the software toward the more convenient branch (point-source or finite-source) in any configurations, thus obtaining another dramatic speed-up far from the caustics. We have adopted pre-calculated tables for the finite-source-single-lens case, which are much faster and have no practical limitations. We have finally described how full light curves for any kind of models can be obtained. For each upgrade we have conducted extensive testing on the whole parameter space, as documented by the figures included in this paper.

Apart from the already mentioned somewhat straightforward extensions to Keplerian orbital motion and arbitrary limb darkening profiles, the new challenges that need to be addressed in view of WFIRST are multiple lenses and astrometry. In this respect, we hope to come back with new versions of our code in the next future.

\section*{Acknowledgements}

\texttt{VBBinaryLensing} has benefited from precious feedback from the community. In particular, we wish to thank Sebastiano Calchi Novati, Giuseppe D'Ago, Radek Poleski, Sedighe Sajadian, Rachel Street, and Wei Zhu for their testing and suggestions.

\appendix

\section{The VBBinaryLensing class}

\texttt{VBBinaryLensing} comes with the classical source file pair \texttt{VBBinaryLens.h} containing the declarations and \texttt{VBBinaryLens.cpp} containing the corresponding implementations. A file named \texttt{instructions.cpp} contains a sample \texttt{main} function with documentation of all features.

In short, the \texttt{VBBinaryLensingLibrary} class contains several functions that perform the microlensing calculations and some variables that can be set by the user to control the optional settings. Here is a list with a short description for each item.

\bigskip

\noindent {\bf Magnification calculations: }
\begin{itemize}
\item \texttt{BinaryMag0}: Magnification of a point-source by a binary lens. There is also a version returning the positions of the images.
\item \texttt{BinaryMag}: Magnification of a uniform brightness-source by a binary lens. There is also a version returning the positions of the images.
\item \texttt{BinaryMagDark}: Magnification of a limb-darkened-source by a binary lens.
\item \texttt{BinaryMagMultiDark}: Magnifications of a limb-darkened-source by a binary lens in different filters with different limb-darkening coefficients. Useful for simultaneous multi-band observations.
\item \texttt{BinaryMag2}: Magnification of a generic source by a binary lens. This function (new in v2.0) implements the tests described in Section \ref{Sec quadrupole}, thus deciding whether to go for point-source or finite-source calculation.
\item \texttt{ESPLMag}: Magnification of a uniform brightness-source by a single lens. This uses the pre-calculated table described in Section \ref{Sec ESPL}.
\item \texttt{LoadESPLTable}: Loads the pre-calculated table for ESPL magnification. It has to be called before the first use of \texttt{ESPLMag} or related functions.
\item \texttt{ESPLMagDark}: Magnification of a limb-darkened-source by a single lens.
\item \texttt{ESPLMag2}: Magnification of a generic source by a single lens. Decides whether to go for point-source or finite-soruce calculation based on the tests in Section \ref{Sec quadrupole}.
\item \texttt{ESPLMagDark}: Magnification of a limb-darkened-source by a single lens.
\end{itemize}

\noindent {\bf Light curve calculations: }
\begin{itemize}
\item \texttt{PSPLLightCurve}: PSPL light curve for a given set of parameters. This and all light curve functions are available for a single epoch or for a full array of observation epochs.
\item \texttt{PSPLLightCurveParallax}: PSPL light curve including parallax.
\item \texttt{SetObjectCoordinates}: Sets the astronomical coordinates of the microlensing target and specifies the path where to look for the position tables of the satellites (if any).
\item \texttt{ESPLLightCurve}: Extended-Source-Point-lens light curve. This uses the \texttt{ESPLMag2} function.
\item \texttt{ESPLLightCurveParallax}: Extended-Source-Point-lens light curve with parallax.
\item \texttt{BinaryLightCurve}: Static binary lens light curve for a given set of parameters. This uses the  \texttt{BinaryMag2} function.
\item \texttt{BinaryLightCurveW}: Static binary lens light curve for a given set of parameters using the center of the caustic of the lens on the right as a reference point for the trajectory.
\item \texttt{BinaryLightCurveParallax}: Binary lens light curve including parallax for a given set of parameters.
\item \texttt{BinaryLightCurveOrbital}: Binary lens light curve including parallax and circular orbital motion for a given set of parameters.
\item \texttt{BinSourceLightCurve}: Light curve for a single lens and a binary source. Sources are treated as point-like.
\item \texttt{BinSourceLightCurveParallax}: Light curve for a single lens and a binary source including parallax.
\item \texttt{BinSourceLightCurveXallarap}: Light curve for a single lens and a binary source including parallax and circular orbital motion.
\end{itemize}

\noindent {\bf Settings: }
\begin{itemize}
\item \texttt{Tol}: Absolute accuracy goal (called $\delta$ in this paper).
\item \texttt{RelTol}: Relative precision goal (called $\epsilon$ in this paper).
\item \texttt{a1}: Linear limb-darkening coefficient used in \texttt{BinaryMag2} and \texttt{ESPLMag2}.
\item \texttt{minannuli}: Minimum number of annuli to calculate for limb-darkening.
\item \texttt{parallaxsystem}: 0 for parallel-perpendicular, 1 for North-East.
\item \texttt{t0\textunderscore par\textunderscore fixed}: Set to 1 if you want to specify a constant $t_{0,par}$.
\item \texttt{t0\textunderscore par}: Reference time for parallax $t_{0,par}$. Only used if \texttt{t0\textunderscore par\textunderscore fixed}$=1$.
\item \texttt{satellite}: Specifies the satellite number for the next calculation ($0$ for observations from the ground).
\end{itemize}

\bsp	
\label{lastpage}
\end{document}